\documentclass[10pt,journal,compsoc]{IEEEtran}

\usepackage{graphicx}
\usepackage[ruled,linesnumbered]{algorithm2e}
\usepackage{caption}
\usepackage{dblfloatfix}    
\usepackage{amsmath}
\usepackage{booktabs}
\usepackage{tabularx}
\usepackage{xcolor}
\newcolumntype{L}{>{\centering\arraybackslash}m{2.5cm}}

\ifCLASSOPTIONcompsoc
  \usepackage[nocompress]{cite}
\else
  \usepackage{cite}
\fi
\ifCLASSINFOpdf
\else
\fi
\begin{document}
%
\title{PowerGAN: A Machine Learning Approach for Power Side-Channel Attack on Compute-in-Memory Accelerators}
%
%
%
%

\author{Ziyu Wang,~\IEEEmembership{Member~IEEE}, 
        Yuting Wu,~\IEEEmembership{}
        Yongmo Park,~\IEEEmembership{}
        Sangmin Yoo,~\IEEEmembership{}
        Xinxin Wang,~\IEEEmembership{}\\
        Jason K. Eshraghian,~\IEEEmembership{Member~IEEE}, 
        and Wei D. Lu*,~\IEEEmembership{Fellow~IEEE}
\IEEEcompsocitemizethanks{\IEEEcompsocthanksitem Z. Wang, Y. Wu, Y. Park, S. Yoo, X. Wang, J. K. Eshraghian and W. D. Lu are with the Department of Electrical Engineering and Computer Science, the University of Michigan, Ann Arbor,
MI, 48109, USA.\protect\\
J. K. Eshraghian is now with Department of Electrical and Computer Engineering, University of California, Santa Cruz, CA 95064, USA.\protect\\
Corresponding Author: Wei D. Lu (wluee@umich.edu)}}
\IEEEtitleabstractindextext{%
\begin{abstract}
  Analog compute-in-memory (CIM) systems are promising for deep neural network (DNN) inference acceleration due to their energy efficiency and high throughput. However, as the use of DNNs expands, protecting user input privacy has become increasingly important. In this paper, we identify a potential security vulnerability wherein an adversary can reconstruct the user’s private input data from a power side-channel attack, under proper data acquisition and pre-processing, even without knowledge of the DNN model. We further demonstrate a machine learning-based attack approach using a generative adversarial network (GAN) to enhance the data reconstruction. Our results show that the attack methodology is effective in reconstructing user inputs from analog CIM accelerator power leakage, even at large noise levels and after countermeasures are applied. Specifically, we demonstrate the efficacy of our approach on an example of U-Net inference chip for brain tumor detection, and show the original magnetic resonance imaging (MRI) medical images can be successfully reconstructed even at a noise-level of 20\% standard deviation of the maximum power signal value. Our study highlights a potential security vulnerability in analog CIM accelerators and raises awareness of using GAN to breach user privacy in such systems.
\end{abstract}

}

\maketitle

\IEEEdisplaynontitleabstractindextext

%
\IEEEpeerreviewmaketitle

\ifCLASSOPTIONcompsoc
\IEEEraisesectionheading{\section{Introduction}\label{sec:introduction}}
\else
\section{Introduction}
\label{sec:introduction}
\fi

%
%
%
%
Machine learning, especially deep neural networks (DNNs), are being used in a broad range of applications, including language processing, computer vision, speech recognition and financial analysis \cite{openai2023gpt}\cite{he2016deep}\cite{variani2014deep}\cite{ozbayoglu2020deep}. However, the state-of-the-art DNN models require significant amounts of data movement between memory and processor during both training and inference, leading to severe memory bottleneck effects \cite{wu2021dynamic} \cite{hung2021challenges} \cite{sebastian2020memory}. Compute-in-memory (CIM) architectures accelerate inference by mapping the pre-trained weights on chip and processing locally in memory, and can thus effectively address the memory bottleneck and result in high energy efficiency and throughput \cite{chen2020survey} \cite{wang2021rram} \cite{chi2016prime}. A typical CIM implementation employs an array of programmable memristive elements, e.g. resistive random-access memory (RRAM), to store the weight values as conductance values and perform vector-matrix multiplication (VMM) through Ohm’s law and Kirchhoff’s current law in place and in parallel \cite{prezioso2015training} \cite{zidan2018future}.

Although DNN models are implemented on a wide range of platforms, from datacenter GPUs to edge devices, the importance of data security cannot be overstated. Attacks on DNN data used in critical applications such as medical diagnosis, autonomous driving, and financial transactions can compromise the user privacy as well as proprietary algorithm information \cite{shokri2017membership}. The main areas of concern in DNN security include model extraction, adversarial attacks, and privacy breaches \cite{shokri2017membership} \cite{hu2020deepsniffer} \cite{gu2014towards}. Model extraction involves extracting the DNN architecture and reconstruct weights to reproduce its functionality, which can be used to breach the intellectual property of DNN design \cite{hu2020deepsniffer}. Adversarial attacks involve manipulating the input data deliberately to make the modification invisible to human but causing misclassification or other unexpected behavior to the DNN model \cite{gu2014towards}. Privacy breaches involve reconstructing sensitive private input or training data by an attacker \cite{shokri2017membership} \cite{wei2018know}.

\begin{figure*}[!t]
  \centering
  \includegraphics[width=\textwidth]{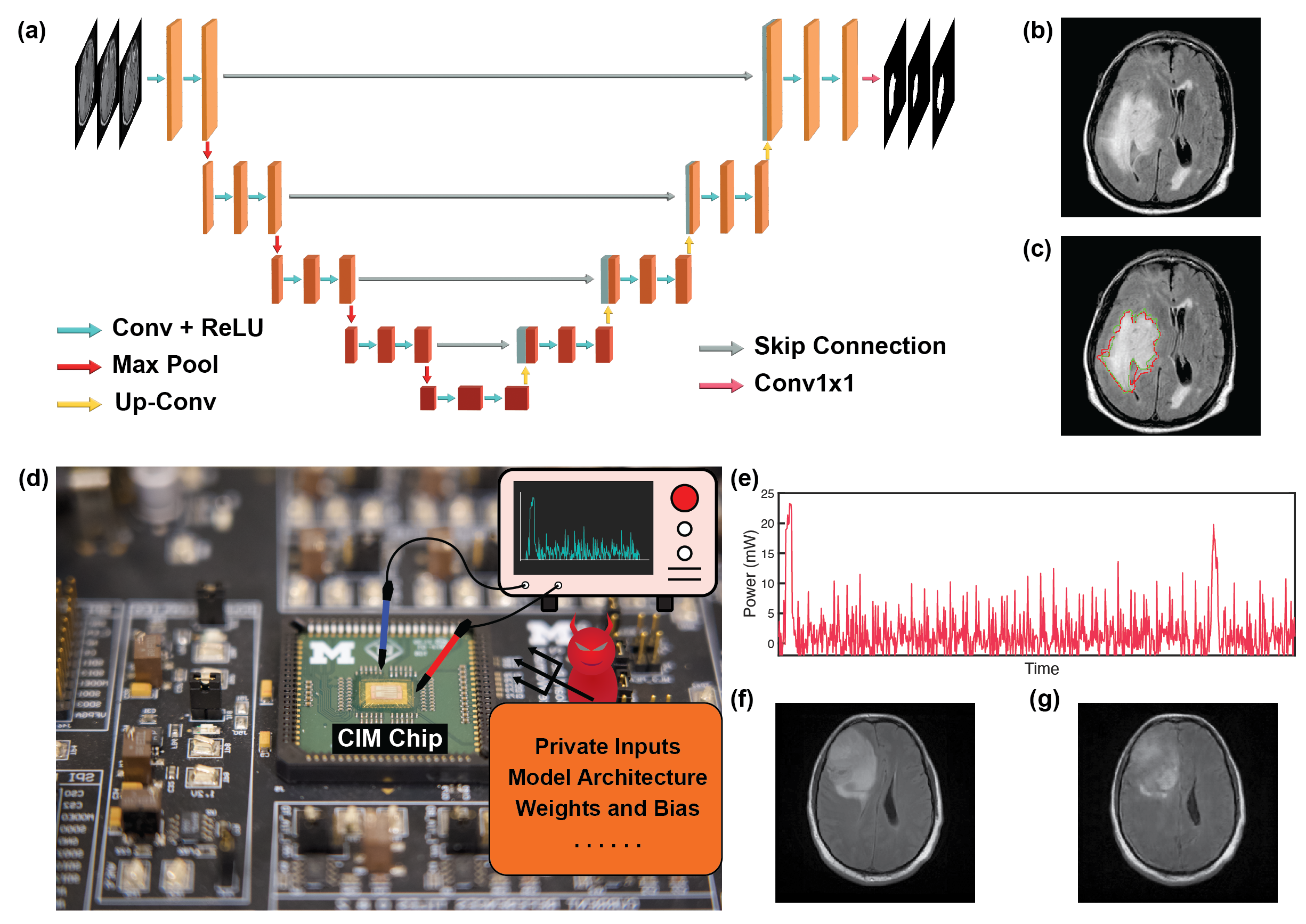}
  \caption{(a) Schematic of the U-Net model architecture for segmentation. It is composed of a down sampling-encoder on the left and an up sampling-decoder on the right forming a U-shaped structure. The skip connections between the encoder and decoder layers are formed by concatenating the feature maps from the encoder layers to the corresponding decoder layers. (b) An example brain MRI image with a tumor. (c) The segmentation result from U-Net. The ground truth and prediction are displayed in green and red, respectively. (d) A CIM chip and the system board. An adversary can potentially perform power side channel attacks through power trace measurements on the chip’s power lines. (e) A representative power trace of a CIM accelerator at inference runtime. (f-g) Examples of a user’s private input sent to the CIM inference accelerator (f) and reconstructed image from power-side channel attacks (g).}
  \label{fig:1}
\end{figure*}

While the abovementioned security attacks on DNNs have been extensively evaluated on systems such as GPUs, CPUs and FPGAs \cite{hu2020deepsniffer} \cite{wei2018know} \cite{xiang2020open} \cite{hua2018reverse} \cite{batina2019csi} \cite{zhang2021stealing}, security and vulnerability analysis of analog CIM accelerators is largely lacking. Recently, Wang et al. \cite{wang2023side} proposed a side-channel attack methodology to extract the DNN architecture from analog RRAM-based CIM systems. Read et al. \cite{read2022method} reverse engineered the weights and biases of DNN models mapped on analog CIM systems. Cherupally et al. \cite{cherupally2022improving} studied adversarial attacks on analog RRAM-based CIM systems. As a potential solution for prevalent low-power edge-based computing, understanding security vulnerabilities of CIM systems including data privacy becomes paramount.

In this study, we propose a machine learning-based approach to reconstruct users’ private input data from power side-channel profiling of CIM systems, without prior knowledge of the DNN model mapped on chip. By scrutinizing the leaked power data features, we identified the security issue of data dependency between power side-channel leakage and private input data. We demonstrated a flexible and automated attack approach based on a generative adversarial network (GAN). The GAN-assisted side-channel attack was shown to be effective even with large noise levels and countermeasure implementations. We evaluated the proposed approach on a human brain magnetic resonance imaging (MRI) dataset \cite{dataset}, which includes detailed features of sulci and gyri in cerebral cortex that can be difficult to reconstruct. In contrast to previous studies on analog CIM system vulnerability, our work highlights a potential significant privacy concern at user’s end.

\section{Background}
\label{sec:Background}
In recent years, RRAM-based CIM systems have been widely studied for DNN inference applications due to their ability to perform in-situ single-step VMM through bit-line current summation \cite{wang2021rram} \cite{prezioso2015training} \cite{zidan2018future}. One area that has greatly benefited from this technology is computer vision, specifically convolution neural network (CNN), which require intensive VMM operations. Moreover, the RRAM-based CIM systems can incorporate transposed convolution for up-sampling in encoder-decoder networks \cite{mao2018lergan} \cite{chen2018regan}. This has been shown to improve efficiency and support various DNN models, as evidenced by several studies \cite{yao2020fully} \cite{wang2019situ} \cite{wu2023bulk} \cite{li2020rram}.

Medical diagnosis, which is a subset of computer vision, involves medical image reconstruction, segmentation, and super-resolution \cite{shen2017deep}, and CIM schemes are promising platforms for medical image processing \cite{zhao2023energy}. The U-Net architecture, one of the most widely used CNN architectures, has been extensively used for image segmentation tasks \cite{ronneberger2015u} \cite{buda2019association}. U-Net, shown in Figure \ref{fig:1}(a), has a “U”-shaped architecture composed of a down-sampling encoder and an up-sampling decoder, which perform feature extraction and reconstruction through convolutional and transposed convolutional layers, respectively. The skip connections in U-Net connect down-sampling and up-sampling paths by concatenating the feature maps to preserve spatial information and improve accuracy. U-Net has been proven effective for multiple medical image segmentation tasks, such as identifying tumors or lesions in MRI scans, even with limited training data. Figure \ref{fig:1}(b) shows an example of an original MRI image used as input to the U-Net studied here, whose outputs are masks of the tumor region. Figure \ref{fig:1}(c) shows the output results of the network, where the green line represents the ground truth and the red line represents the network prediction results.

\subsection{Side-Channel Attacks}

CIM systems use pre-trained models for inference acceleration on chip. Any potential security threat can be exploited by a malicious adversary who could launch a side-channel attack on the system through the side-channel leakage. Side-channel attacks aim at extracting private data from a hardware system by measuring and analyzing physical parameters during execution \cite{kocher1996timing} \cite{kocher1999differential}. Such parameters include supply power, execution time, and electromagnetic emission. Attackers can reverse engineer the sensitive data or architectural information by deliberately measuring and analyzing the side-channel dissipation of the chip. Side-channel data acquisition can be either invasive or non-invasive, depending on whether the chip needs to be decapsulated \cite{hutle2015resilience} \cite{fan2010state}. Both can be considered to measure the data dependent signals leaked from RRAM tiles. As RRAM devices are fabricated in the back-end-of-line, invasively etching away the passivation layers of CIM macros can provide immediate access to the top-level metal lines of each tile for potential probing. The adversary can then probe the power lines directly and extract the power data using an oscilloscope, as shown in Figure \ref{fig:1}(d). Non-invasive techniques include the use of electromagnetic probes to measure the electromagnetic emission of RRAM tiles, taking advantage of the spatial locality of model mapping \cite{funato2006magnetic} \cite{chou2013space} \cite{peng2019pair}. The side-channel attack is performed on measured leaked traces, with one such example shown in Figure \ref{fig:1}(e), where the power trace of a CIM module at inference runtime is shown. In this work, the goal is to reconstruct the user’s private input data from power traces measurements. Figure \ref{fig:1}(f) and (g) show an example of the original private input and the reconstruction result using the proposed approach, respectively.

Accurately measuring side-channel leakage signals requires sophisticated data acquisition scheme design and high-precision measurement equipment. To study the vulnerability of the chip design and secure it iteratively, it is more practical to simulate the attack scenarios using real device measurement data, followed by redesigning the chip with security considerations before massive production, especially for the emerging CIM systems. In our study, we simulate an CIM system designed with TSMC 28 nm technology, with details described in \cite{wang2023side}.

A general assumption for side-channel attacks is that the attacker already knows the hardware architecture and implementation of the CIM chip, which in this case includes the RRAM array size and a mixed-signal interface design that includes analog-to-digital converters (ADCs). However, the attacker has no knowledge of the neural network mapped on chip. Although the inference tasks of the RRAM-based CIM system can be pipelined to improve the throughput, the attackers have full control over the input and output sequence, allowing them to halt the next input until the completion of the current one. This enables the attacker to gain more accurate power measurements without blurring valuable power signals. The finest grain of the chip that an attacker can access is a tile, and they have no access to individual RRAM cells.

\section{CIM System for DNN Inference Acceleration}
\begin{figure*}[!t]
  \centering
  \includegraphics[width=\textwidth]{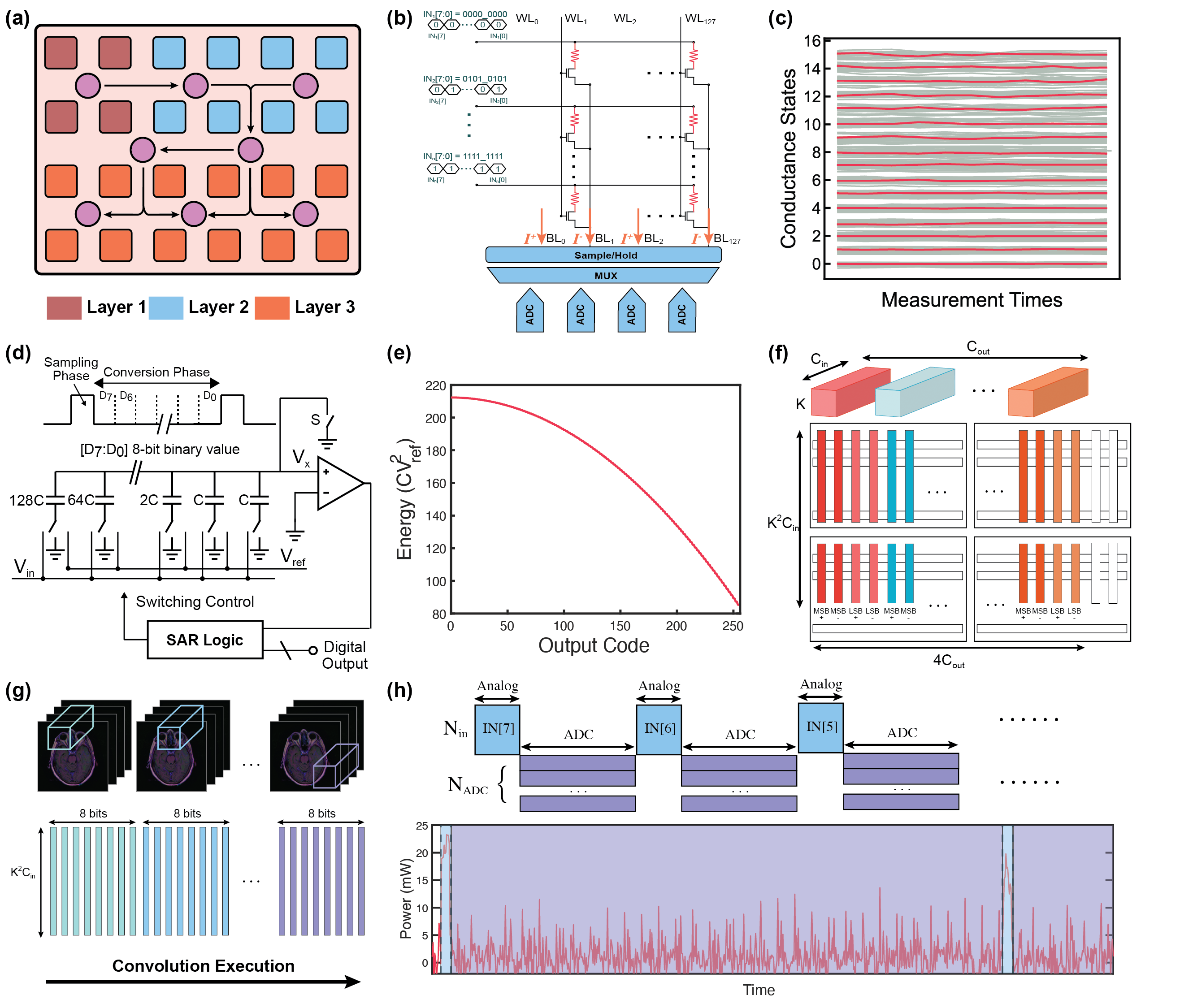}
  \caption{Schematic diagrams of (a) DNN mapping on a tiled CIM accelerator. (b) A CIM tile with bit-serial inputs and shared ADC. (c) Readout measurements of the devices’ state after the devices are programed to 16 conductance levels. (d) Schematic of an 8-bit charge redistribution SAR ADC with timing diagram is inset at the top. (e) Switching energy of the SAR ADC with respect to the ADC output code. (f) Weight mapping scheme when 4 ADCs are shared by 128 columns in 1 tile. (g) 8-bit bit-serial input activation mapping at inference runtime. (h) Timing diagram of CIM accelerator at inference runtime, with bit-serial inputs applied to the array (blue) followed by analog-to-digital conversion (purple). The execution sequence of analog array and ADC is marked with corresponding colors in the power trace below.}
  \label{fig:2}
\end{figure*}

\subsection{RRAM Array and Device}
In analog RRAM-based CIM system, pretrained DNN weight matrices are mapped on RRAM macros in a tiled architecture, as shown in Figure \ref{fig:2}(a). VMMs are performed in each tile, and the outputs are accumulated from tiles belonging to the same layer, followed by neuron function conversion before being passed to the next layer through on-chip routers. Detailed schematic of an analog RRAM-based CIM tile is shown in Figure \ref{fig:2}(b). The pretrained weights are mapped to device conductance values according to the device conductance range. RRAM devices can support multi-bit mapping as they offer multiple conductance levels due to the gradual modulation of conductive paths \cite{wu2022demonstration} \cite{du2015biorealistic} \cite{sun2019modulating}. Our experiments are based on devices with 16 conductance levels, i.e. 4-bit precision, as shown in Figure \ref{fig:2}(c). Bit-serial input is used in this study. The outputs of VMMs are returned as bit-line currents, subsequently sampled by the peripheral readout circuitry and converted to binary digital values by ADC for further downstream processing and communication. 

\subsection{Mixed-Signal Interface}
Figure \ref{fig:2}(d) depicts the schematic of an 8-bit charge redistribution successive approximation registers (SAR) ADC utilized in the mixed-signal interface. A SAR ADC is composed of three parts: a comparator, a SAR logic controller, and a capacitive digital-to-analog converter (DAC). Of these, the DAC consumes the majority of the overall power dissipation \cite{saberi2011analysis}. Compared with other approaches, the SAR ADC has advantages in process scaling, power efficiency, high precision and conversion speed \cite{harpe2016successive}. It converts the analog inputs by switching each capacitor in the DAC from $V_{in}$ to $V_{ref}$ one by one to perform a binary successive approximation, as shown in the timing diagram inset in Figure \ref{fig:2}(d). The conversion takes eight consecutive steps, and the energy consumption of the n-th step can be calculated from the change in $V_x$, and the capacitance connected to $V_{ref}$ using Equation \ref{eq:1},

\begin{equation}
   E=
   \begin{cases}
   -C_{1}V_{ref}(\Delta V_{x} - V_{ref}), & n = 1\\ 
   -V_{ref}(\Delta V_{x} \sum_{i=1}^{n-1}C_{i}D_{i} + C_{n}(\Delta V_{x} - V_{ref})), & n \neq 1
   \end{cases}
  \label{eq:1}
\end{equation}

\noindent where $D_{i}$ is the $i$~th MSB output code. When $D_{i} = 1$, the $i$-th switch connects to $V_{ref}$, otherwise, it connects to ground.

The energy during the eight DAC switching steps dominates the total analog-digital conversion energy consumption. Figure \ref{fig:2}(e) shows the normalized total DAC switching energy, and it exhibits a clear data pattern dependency \cite{ginsburg2005energy} \cite{hariprasath2010merged}.

\subsection{Mapping and Inference Execution}
For inference tasks, pretrained weights are mapped across RRAM tiles, and remain fixed during operations. To balance accuracy and inference efficiency, the weights and input activations are typically quantized to 8 bits, which has been shown to cause minor accuracy losses, especially when coupled with quantization-aware training techniques \cite{wu2023bulk}. Since a single RRAM cell considered here only offers 4-bit storage, multiple cells are used to store one weight value. Figure \ref{fig:2}(f) illustrates the mapping scheme of 8-bit weights from a convolution layer to 4-bit devices in Figure \ref{fig:2}(c). Each convolution kernel is flattened to a vector, and the positive and negative values are separated into two columns. The weights are then quantized to 8-bit precision and split into two 4-bit numbers based on their significance (i.e., 4 MSBs and 4 LSBs). In this case, the total number of crossbar columns for mapping is four times the output channel size, with two for polarity and two for significance. The number of crossbar rows is the flattened kernel size, $K^2C_{in}$, where K is the kernel dimension and $C_{in}$ in the input feature depth.

The output of the convolutional layer is computed by sliding the convolution window through the input feature maps, as illustrated in Figure \ref{fig:2}(g). The inputs are flattened to vectors to match the kernels, and outputs from all kernels are computed simultaneously through CIM tiles. Each 8-bit input is mapped to bit-serial represented voltage pulses, and it takes eight steps to apply them on the crossbar array. The timing diagram of CIM tiles during computation is shown in Figure \ref{fig:2}(h). An input bit is applied to the array, and the resulting output current from bit-lines is sampled and held by the peripheral circuit. For area and power efficiency considerations, chip designers often implement an ADC-sharing scheme instead of assigning each column with an ADC. The analog outputs from the shared columns are consecutively converted by the ADC through the multiplexer (MUX). In our experiment, the number of ADCs per array is set to 4. The power trace shown in Figure \ref{fig:2}(h) is simulated for the abovementioned U-Net and hardware configurations through the simulation flow described in \cite{wang2023side}. The blue regions represent the analog computation cycle, and the purple regions represent ADC conversion cycle. Due to the time required to read a stable output from the analog computation, the analog execution can be readily identified from the ADC execution power traces even with high measurement noise.

\begin{figure*}[!t]
  \centering
  \includegraphics[width=\textwidth]{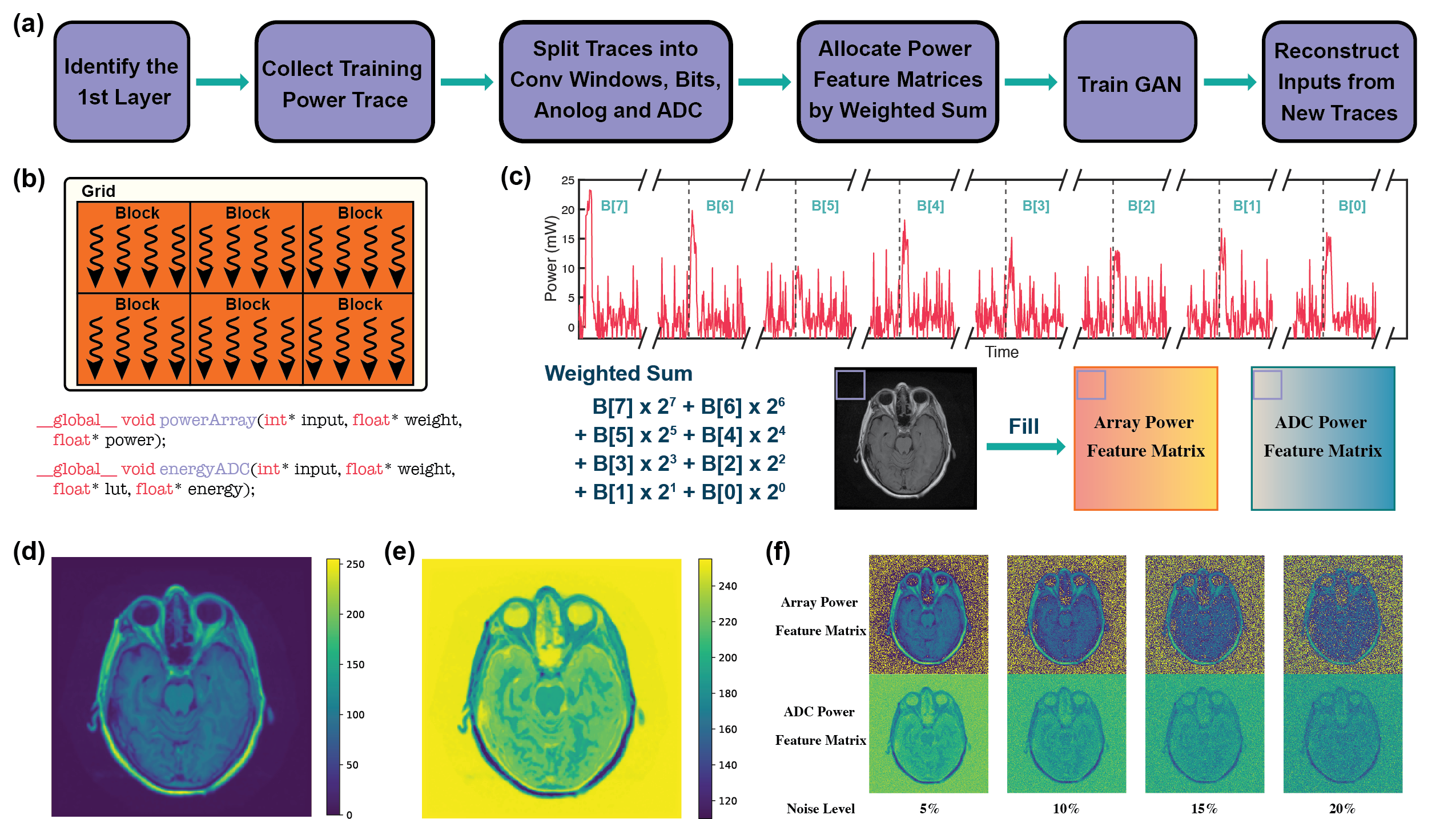}
  \caption{(a) Overview of power side-channel attack flow for private input reconstruction. (b) Block diagram of CUDA SIMT execution for fast simulation of the crossbar array analog computing power and ADC switching energy. (c) Data preprocessing steps after power trace acquisition: First, the power trace in each sliding window is broken into single bits in bit-serial input format, where each bit execution contains an analog computing and an analog-to-digital conversion phase. Second, the power consumption of each sliding window is weighted based on the bit significancy. Third, the weighted sum results are recorded in the two power feature matrices, corresponding to array analog computing and ADC conversion, respectively. (d) Array power feature matrix, and (e) ADC power feature matrix obtained in an ideal simulation without adding noise, normalized into 8-bit unsigned integer range. They represent the extracted input information leakages. (f) Power feature matrices with different levels of noise injection. The noise level corresponds to the ratio of the standard deviation to the maximum value in the power feature matrix.}
  \label{fig:3}
\end{figure*}

\section{Side-Channel Leakage in CIM Systems}
\subsection{Attack Flow and Power Modeling}
Figure \ref{fig:3}(a) illustrates the attack flow aimed at reconstructing private input data from the power side-channel leakage. The attack targets the first layer of the DNN model since it is the closest layer to the input port and directly executes the input data. The methodologies proposed in \cite{wang2023side} can be used to extract the property of the first layer, i.e., structure of the convolutional layer and the associated CIM tiles used to execute the layer. As DNN models are trained on specific tasks, an adversary can collect power traces using similar input data and learn the dependency between them. In this case, the adversary can feed other MRI images and collect their own power traces to train an attack model before attacking unknow inputs. Once power traces are allocated, power feature extraction and data preprocessing are required to find the correlation between input and leakage. We refer to the data after preprocessing as power feature matrices, which can be utilized to train a machine learning model for input data reconstruction. In our study, we employed a GAN for reconstruction since it shows good noise tolerance and can overcome noise-injection countermeasures.

The power trace simulator proposed in \cite{wang2023side} provides valuable insights into dynamic power and timing data. However, processing every single data point in a large dataset is impractical due to the time-consuming nature of simulating power dissipation at sub-nanosecond-level precision, as well as the generation of tremendous data files. To address this, we developed a fast power feature simulator based on NVIDIA Compute Unified Device Architecture (CUDA), leveraging modern GPU’s single instruction multiple threads (SIMT) to process every bit input in the bit-serial fashion, as shown in Figure \ref{fig:3}(b). The CUDA kernel functions simulate power dissipation and total energy of RRAM arrays and ADCs at each execution, respectively. Both kernels take bit-serial inputs and conductance weight matrices as inputs. The kernel function for array power simulation computes the power using applied input voltage and device conductance values. The kernel function for ADC is more complex. First, we compute the switching energy of each output code (Figure \ref{fig:2}(e)) and store them in a lookup table (LUT). Then, the kernel function computes the analog current of each bit-line, followed by scaling it into the proper range to index the ADC energy from the LUT. Due to ADC sharing, the ADC energy in one execution is the sum of four ADCs.

To convert all outputs in an array with 128 columns and 4 ADCs requires 32 executions. However, the ADC traces are similar to each other, making it impractical to distinguish every execution with measurement noise, as shown in the noisy trace Figure \ref{fig:3}(c). Therefore, we treat the energy of all 32 ADC executions as a whole for further data preprocessing.

\subsection{Data Preprocessing and Leakage}
When an input image is processed by the model stored on the CIM chip, the convolution window in the first convolution layer slides through the entire input image. Here we define two power feature matrices, corresponding to the analog array computation and ADC conversion energy when performing the convolution operation, respectively. The power feature matrices will have the same size as the output feature map of the first convolution layer. An entry in a power feature matrix corresponds to the collected power information during the convolution operation at the corresponding position in the input feature map. Within each convolution window, the input data are reshaped and scaled into 8 bits before being applied to the crossbar array. Processing an input bit in CIM can be divided in two steps: analog array computation and analog-to-digital conversion, as shown in Figure \ref{fig:3}(c). Because the eight input bits are of varying significancy, the input-dependent power leakage at each bit computation should not be treated equally. To recover the input-dependent power data, we used a weighted sum approach for the array power and ADC energy according to the bit significance, as shown in Equation \ref{eq:2} and \ref{eq:3}.

\begin{equation}
   P_{array}=\sum_{i=0}^7{P_{array}[i] \times 2^i}
  \label{eq:2}
\end{equation}

\begin{equation}
    E_{ADC}=\sum_{i=0}^7{E_{ADC}[i] \times 2^i}
  \label{eq:3}
\end{equation}

Once the weighted sum results of each convolution window have been computed, they are then populated into the two power feature matrices. Figure \ref{fig:3}(d) and (e) show examples of the obtained array power feature matrix and the ADC power feature matrix, respectively, without considering noise in the power data measurement. The power side-channel leakage exhibits a strong dependency on the input data and reveals a security risk. 

Both the array power dissipation and ADC energy consumption are functions of the inputs and weights, which explains the strong dependency between the power side-channel leakage and input data. During inference, the weights are held constant leading to a linear relationship between the power feature matrices and the original input, as shown in Equation \ref{eq:4},

\begin{equation}
    y_{i, j}=\mathcal{F}\left(\left[
    \begin{matrix}
    x_{i-r,\ j-r} & \cdots & x_{i-r,\ j+r} \\
    \vdots & \ddots & \vdots \\
    x_{i+r,\ j-r} & \cdots & x_{i+r,\ j+r}\\
    \end{matrix}
    \right]\right) 
  \label{eq:4}
\end{equation}

where i, j are indices of the entry from power feature matrices, r is the radius of the convolution window, x is the the input data and y is the corresponding entry in the power feature matrices. The function $\mathcal{F}$ is used to convert the input data into power feature data after weighted sum. Regardless of whether $\mathcal{F}$ represents array or ADC, it is always a one-to-one projection. As a result, the input information is preserved in the power feature matrices and leading to a severe security issue.

Additionally, as shown in Figure \ref{fig:3}(d) and (e), the array power feature matrix contains more detailed information than the ADC power feature matrix. This is because each entry in the ADC power feature matrix represents the total energy of 32 executions, with each execution involving four ADCs operating together. Compared with the array computation power feature matrix which entries indicating single execution, the ADC power feature matrix produces coarser granularity. Furthermore, from the weight mapping scheme in Figure \ref{fig:2}(f), the energy consumption of the four ADCs are associated with a weight value in four representations of MSB+, MSB-, LSB+ and LSB-. Since the LSBs and MSBs have different impacts on the weight value, the ADC energy output is not a direct linear transformation of the input data, where the array power feature matrix corresponds to a linear transformation of the input data (directly proportional to the product of the input and the weight matrix).

\subsection{Noise and Countermeasures}
The power side-channel attack approach mentioned above is capable of reverse engineering private inputs. However, for a real-world applications non-ideal effects such as noise during data acquisition must be considered. Noise can originate from multiple sources of the chip, including thermal noise and human-made noise \cite{rohrer1971computationally} \cite{schreier2005design} \cite{sheikholeslami2020power}. Thermal noise is a physical phenomenon that cannot be eliminated. Thermal noise can be found on RRAM devices. Although ideal capacitors have no thermal noise, when they are coupled with other components in the circuit, there will be a combination of kTC noise. Other noise from power lines or measurement equipment may have higher noise power and will lower the signal-to-noise ratio when the adversary measures the power side-channel leakage. The power feature data, accounting for the presence of noise, are obscured by a noise term N, which can be mathematically described by Equation \ref{eq:5}.

\begin{equation}
    y_{i, j}=\mathcal{F}\left(\left[
    \begin{matrix}
    x_{i-r,\ j-r} & \cdots & x_{i-r,\ j+r} \\
    \vdots & \ddots & \vdots \\
    x_{i+r,\ j-r} & \cdots & x_{i+r,\ j+r}\\
    \end{matrix}
    \right]\right) + \mathcal{N}\left(\cdots\right)
  \label{eq:5}
\end{equation}

Noise can also be used to protect the system. One common countermeasure to mitigate power analysis side-channel attacks is noise injection. Noise injection works by adding a random noise signal to the original signal to mask the correlation between the leaked information and the secret information \cite{spreitzer2017systematic} \cite{das2017high}. The noise signal is designed to be random and uncorrelated with the original signals, so statistical analysis methods for denoising may not be valid anymore.

\begin{figure*}[!t]
  \centering
  \includegraphics[width=\textwidth]{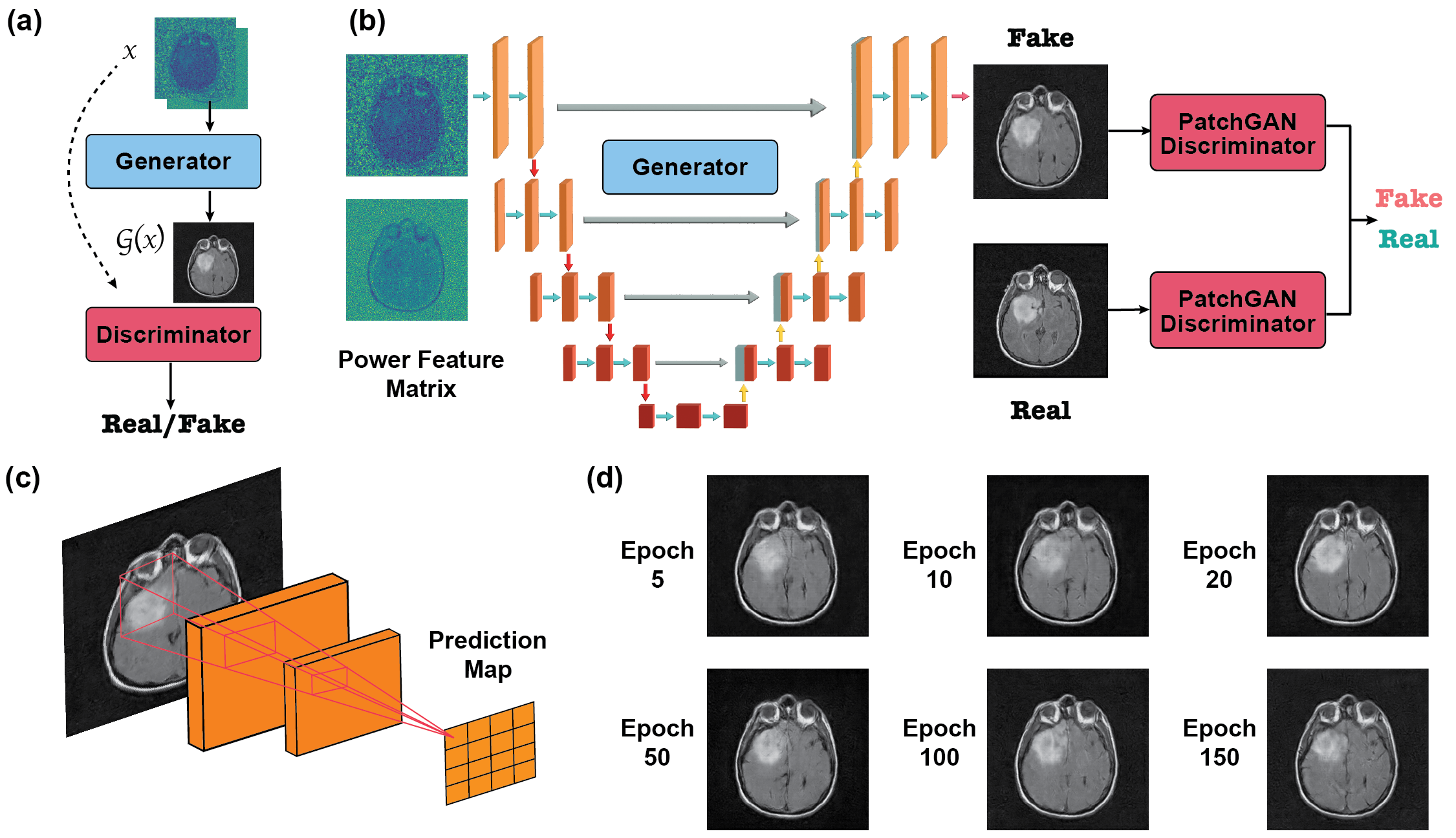}
  \caption{(a) Schematic of the conditional GAN model architecture. (b) Schematic of the pix2pix GAN architecture for reconstructing image from noisy power traces. The power feature matrices are concatenated as the input to the generator, which is based on a U-Net architecture. The generated image, along with the original image and the power feature matrices, are fed to the PatchGAN discriminator for differentiation. (c) PatchGAN architecture, which outputs a 2D score matrix where each element corresponds to a local patch of the input image. (d) Evolution of the image reconstruction from power feature matrices over 150 training epochs, with a noise level of 20\%.}
  \label{fig:4}
\end{figure*}

In Figure \ref{fig:3}(f), we simulate different noise levels from 5\% to 20\% at side-channel data acquisition. The percentage levels here are defined as the ratio of the noise’s standard deviation to the maximum measured signal in the power feature. As the noise level increases, details of the cerebrum region in the power feature matrices are lost. It is noteworthy that when the noise level reaches 20\%, the cerebrum region was effectively masked from the power feature matrices, and only noise is circled by the skull. Hence, noise injection is a powerful countermeasure to mitigate the side-channel leakage in the CIM system. To deal with noise injection, at the adversary end, a more effective attack approach is required.

\section{Machine Learning Assisted Side-Channel Attack}
\subsection{Generative Adversarial Network for Side-Channel Attac}
Adversaries often attempt to design elaborate denoising schemes to stripe the noise signal and expose the valuable original signal \cite{wei2018know}. However, using conventional denoising techniques requires considerable effort in denoising design at both the hardware and algorithm levels. The adversary needs to specify the noise frequency and apply a low-pass filter to cut off high-frequency noise during measurement. Then, the adversary needs to identify the working spectrum of the CIM system, and recover the distortion induced by the power measurement circuit. To establish the relationship between the restored power curve and the original data, the adversary is required to conduct circuit analysis of the power data at each execution frame, which can be inefficient when dealing with large volumes of input. Therefore, a flexible attacking approach with noise tolerance is essential for efficient side-channel attacks and for prompting to designing more secure and reliable CIM systems.

Machine learning approaches have recently been explored for side-channel attacks since they can be highly automated and scalable, allowing attackers to extract sensitive information with minimal human intervention from large volumes of data \cite{picek2023sok} \cite{kubota2021deep} \cite{hettwer2020applications}. Machine learning-based side-channel attacks involve a training phase and an attack phase. The training phase can be controlled by the adversary by building a leakage model from the trace collected earlier using known input data, which alleviates the requirement of collecting sufficient traces from limited resources. The adversary can design a system-specific model to prompt a side-channel attack and recover the target data from newly measured trace during the attack phase.

GANs are a type of DNN architecture consisting of two neural networks for generating synthetic data: a generator and a discriminator \cite{goodfellow2020generative}. In conventional GANs, the generator takes random noise as input and tries to generate data to mimic the training data, while the discriminator takes both the real data and generated data and tries to distinguish between them. During training, the two networks are trained together in a min-max game, where the generator tries to produce data that can fool the discriminator, and the discriminator tries to evaluate the authenticity of the generated data. Unlike conventional GANs, which take random noise as input, conditional GANs (cGANs) take extra information as input, such as image features or text description \cite{mirza2014conditional} \cite{reed2016generative} \cite{isola2017image} \cite{zhu2017unpaired}, as shown in Figure \ref{fig:4}(a). The target of cGANs is to generate more controlled outputs belonging to a certain category or containing certain desired features. cGANs have many potential applications in various fields such as image-to-image translation, text-to-image synthesis, and style transfer. In a side-channel attack, the adversary’s goal is to train a neural network to reconstruct the input information from side-channel leakage, making cGANs a good fit for this task.

Noise is required during training of both conventional GANs and cGANs because it provides the generator with a source of randomness, allowing it to produce diverse and realistic data and improve the model performance. Consequently, GANs are excellent candidates for coping with noise (and leveraging the noise) during side-channel data measurement, enabling the elimination of complex conventional denoising schemes for data acquisition. GAN-based DNNs can make side-channel attacks more flexible and enhance attack success rate.

\subsection{Experiment Setup}
We note the reconstruction of the original image from power feature matrices is analogous to image-to-image translation. In this study, we adopted the pip2pix cGAN \cite{isola2017image} architecture to reconstruct the input images from side-channel leakage. Pix2pix is a specialized version of cGANs designed to map an input image from one domain to an output image in another domain. 

The pix2pix architecture used in our experiment is shown in Figure \ref{fig:4}(b). The generator is a standard U-Net architecture, similar to the one discussed in Figure \ref{fig:1}(a) for medical image segmentation. The array power feature matrix and ADC power feature matrix, shown above and below in the left of Figure \ref{fig:4}(a), are concatenated in the channel direction before being fed into the neural network. The discriminator is a patch-based CNN called PatchGAN. Two discriminator networks are used in the discrimination phase, by analyzing pairs of images, a real MRI image and a fake image generated by the generator, along with the stacked power feature matrices as input, and outputs a patch-level prediction. The architecture of the PatchGAN discriminator is shown in Figure \ref{fig:4}(c). For a input image, the PatchGAN takes a patch of the image as input and outputs a matrix of values, where each value in the matrix indicates the probability that the corresponding patch in the input image is real. For the $256 \times 256$ images, patch size of $70 \times 70$ can achieve realistic reconstruction results. In this case, the prediction map size is $4 \times 4$. By computing the average of the values in the prediction map, the PatchGAN can measure the overall realism of the generated images.

PatchGAN has several advantages in reconstructing the lost details in the power feature matrices. Firstly, it encourages the generator to produce more detailed and high-frequency information in the output, as it has to fool the discriminator at the smaller patch level (vs the larger image level). Secondly, the discriminator can provide more fine-grained feedback to the generator since it evaluates images at the patch level. Thirdly, by using smaller patches rather than the entire image, PatchGAN can capture local image features. 

The pix2pix cGAN model is trained on the brain MRI dataset with power feature matrices with size of $256 \times 256$. A total of 3143 MIR image data are used in 200 training epochs with batch size of 1. The data are split into train, validation, and test sets randomly.

\begin{figure*}[!t]
  \centering
  \includegraphics[width=\textwidth]{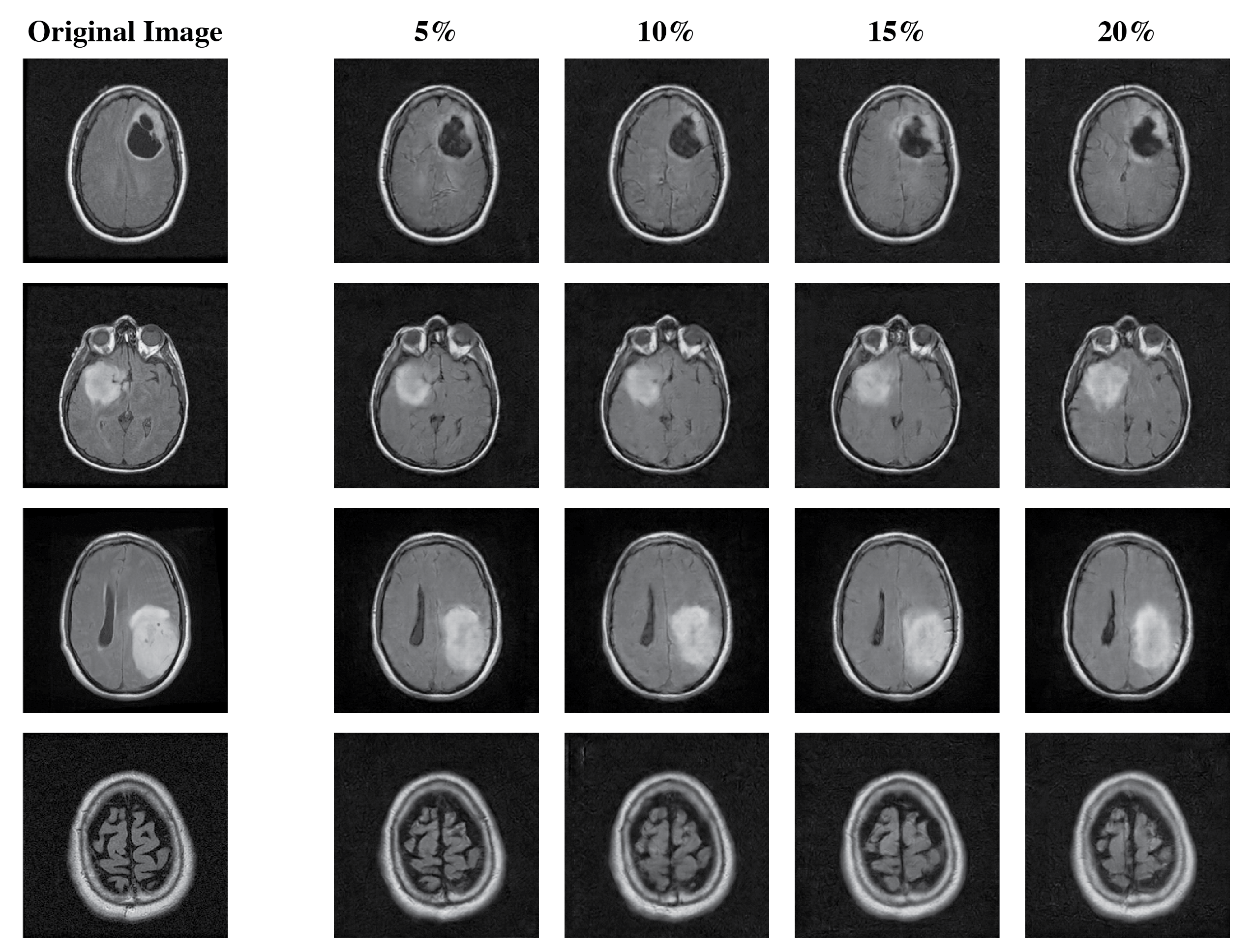}
  \caption{Image reconstruction of four representative brain MRI images. The left column shows the original MRI images, and the right columns show reconstruction results at different injected noise levels.}
  \label{fig:5}
\end{figure*}

\subsection{Results and Discussion}
Figure \ref{fig:4}(d) shows the image reconstruction results as a function of training epochs. The test image used for reconstruction was injected with a 20\% noise level in the power feature matrices. The model is capable of generating a high-level brain structure, including a highlighted tumor region, in just 5 epochs. However, some tiling artifacts can be observed in the reconstructed images when the epoch number is below 20, and the detailed information of brain lobes has not been fully reconstructed. The model is fine-tuned in the subsequent epochs to achieve more precise reconstructions of the MRI images. Human brains are characterized by a folded cerebral cortex, which exhibits diverse details across different MRI images, which
makes it hard to reconstruct all detailed information. As the epoch number increases, the model was able to reconstruct more details of sulci and gyri in the cerebral cortex, along with more precise locations of the tumor region. 

Figure \ref{fig:5} shows reconstruction results for brain MRI images corresponding to different horizontal sections, tumor types and brain lobes structures, with varying levels of noise. The original images are shown on the left, and the reconstructed counterparts with different noise levels are shown on the right. All the reconstructed images demonstrate accurately restored large-scale structures, including tumor type, tumor region, and brain lobes structures. Furthermore, local details such as sulci and gyri in the cerebral cortex are also well preserved in the reconstruction results, and closely resemble the original MRI images. It should be noted that as the noise level increases, the quality of the reconstructed images slighted decreases, leading to slightly blurred images and missing or adding sulci and gyri. However, most of significant information from the original images remains intact, even at high noise levels of up to 20\%. Consequently, the GAN-assisted side-channel attack is effective in defeating this level of noise injection countermeasures in CIM systems.

\subsection{Conclusion}
In this study, we analyzed power side-channel attacks on CIM accelerators, and show carefully designed side-channel attacks can reverse engineer private input data from the user without any prior knowledge of the DNN model used on the chip, thus revealing a potential significant security vulnerability. We propose an automated input reconstruction scheme based on pix2pix cGAN for input image reconstruction from power side-channel leakage, and demonstrate the effectiveness of the proposed attack method on a brain MRI dataset. Our experiments show the power side-channel attack can tolerate a high noise level in power data acquisition, and defeat conventional noise-injection countermeasures. This work highlights a critical vulnerability in CIM systems and underscores the need for greater attention to security considerations in CIM architecture design.

\ifCLASSOPTIONcompsoc
  \section*{Acknowledgments}
\else
  \section*{Acknowledgment}
\fi

This work was support in part by SRC and DARPA through the Applications Driving Architectures (ADA) Research Center, and by the National Science Foundation through Award CCF-1900675.

\ifCLASSOPTIONcaptionsoff
  \newpage
\fi

\bibliographystyle{IEEEtran}
\bibliography{ref}

\end{document}